\documentclass[12pt]{article}
\usepackage{graphicx}
\usepackage{amssymb}
\newcommand{\beq}{\begin{equation}}
\newcommand{\eeq}{\end{equation}}
\newcommand{\beqn}{\begin{eqnarray}}
\newcommand{\eeqn}{\end{eqnarray}}

\newcommand{\bv}{\mathbf{v}}

\newcommand{\ep}{\mbox{${\varepsilon}$}}
\newcommand{\gsim}{\mbox{$>$\hspace{-0.8em}\raisebox{-0.4em}{$\sim$}}}

\begin{document}

\begin{center}
{\Large \bf  Capture of dark matter by the Solar System.\\
\vspace{3mm} Simple estimates}
\end{center}

\vspace{1cm}

\begin{center}
I.B.~Khriplovich\footnote{khriplovich@inp.nsk.su}\\
Budker Institute of Nuclear Physics\\
630090 Novosibirsk, Russia,\\
\end{center}

\vspace{1cm}

\begin{abstract}
We consider the capture of galactic dark matter by the Solar
System, due to the gravitational three-body interaction of the
Sun, a planet, and a dark matter particle. Simple estimates are
presented for the capture cross-section, as well as for density
and velocity distribution of captured dark matter particles close
to the Earth.
\end{abstract}

\vspace{1cm}

\section{Introduction}

The density of dark matter (dm) in our Galaxy is (see, e.g., \cite{ber})
\begin{equation} \label{rho}
\rho_g \simeq 4 \cdot 10^{-25}\;\rm{g/cm}^3\,.
\end{equation}
However, only upper limits on the level of $10^{-19}\;
\rm{g/cm^3}$ (see \cite{khp, kh}) are known for the density
of dark matter particles (dmp) in the Solar System (SS). Besides, even
these limits are derived under the quite strong assumption that
the distribution of dm density in the SS is spherically-symmetric
with respect to the Sun. Meanwhile, information on dm density in SS
is very important, in particular for the experiments aimed at the
detection of dark matter.

The capture of dark matter by the SS was addressed previously in
[4 -- 8]. In particular, in~\cite{khs} the total mass of the captured dark matter was
estimated analytically. In the present note the analytical
estimates are given for the capture cross-section, as well as for
the density and velocity distribution of captured dm close to the
Earth.

Of course, a particle cannot be captured by the Sun alone. The
interaction with a planet is necessary for it, i.e. this is essentially
a three-body (the Sun, planet and dmp) problem. Obviously, the
capture is dominated by the particles with orbits close to
parabolic ones with respect to the Sun; besides, the distances
between their perihelia and the Sun should be comparable with the
radius of the planet orbit $r_p\,$. Just the trajectories of these
particles are most sensitive to the attractive perturbation by the
planet.

The capture can be effectively described by the so-called restricted
three-body problem (see, e.g., \cite{sz}). In this approach the
interaction between two heavy bodies (the Sun and a planet in our case)
is treated exactly. As exactly is treated the motion of the third,
light body (a dmp in our case) in the gravitational field of the two
heavy ones. One neglects however the back reaction of a light particle
upon the motion of the two heavy bodies. Obviously, this
approximation is fully legitimate for our purpose.

Still, the restricted three-body problem is rather complicated,
its solution requires both subtle analytical treatment and serious
numerical calculations (see, for instance, \cite{pe}). Under
certain conditions the dynamics of light particle becomes chaotic.
The "chaotic" effects are extremely important for the problem.
However their quantitative investigation is quite complicated and
remains beyond the scope of the present note. We confine here
instead to simple estimates which could be also of a methodological
interest by themselves. On the other hand, thus derived results
for the total mass and density of the captured dark matter constitute at least an upper
limit for their true value. As to the velocity distribution of dmp's
given here, together with the mentioned result for
the dark matter density, it could be possibly of some practical interest for
planning the experimental searches for dm.

\section{Total mass of dark matter captured by the Earth}
The Solar System is immersed in the halo of dark matter and moves
together with it around the center of our Galaxy. To simplify the
estimates, we assume that the Sun is at rest with respect to the
halo. The dark matter particles in the halo are assumed also to
have in the reference frame, comoving with the halo, the Maxwell
distribution (see \cite{ag}):
\begin{equation}
\label{ma} f(v)\,dv = \sqrt{\frac{54}{\pi}}\,\frac{v^2 dv}{u^3}
\exp{\left(-\frac{3}{2}\frac{v^2}{u^2}\right)}\,,
\end{equation}
with the local rms velocity $u \simeq 220$ km/s. Let us note that
the velocities $v$ discussed in this section are the asymptotic
ones, they refer to large distances from the Sun, so that their
values start at $v=0$ and formally extend to $\infty$.

The amount of dm captured by the SS can be found by means of
simple estimates\footnote{These estimates were given previously in
\cite{khs}. Here we repeat them, as well as results (\ref{de1}),
(\ref{E}), (\ref{td}) (see below), since they are essential for
the present discussions.}. The total mass captured by the Sun (its
mass is $M$) together with a planet with the mass $m_p$, during
the lifetime
\begin{equation} \label{T}
T \simeq 4.5 \cdot 10^9 \;\rm{years} \simeq   10^{17}\; \rm{s}
\end{equation}
of the SS, can be written as follows:
\begin{equation}\label{de}
\mu_p = \rho_g\, T < v\,\sigma>\,;
\end{equation}
here $\sigma$ is the capture cross-section. The product $\sigma v$
is averaged over distribution (\ref{ma}); with all typical
velocities in the SS much smaller than $u$, this distribution
simplifies to
\begin{equation} \label{rj}
f(v)\,dv = \sqrt{\frac{54}{\pi}}\,\frac{v^2 dv}{u^3}\,.
\end{equation}

To estimate the average value $<v \,\sigma>$, we resort to
dimensional arguments, supplemented by two rather obvious physical
requirements: the masses $m_p$ and $M$ of the two heavy components
of our restricted three-body problem should enter the result
symmetrically, and the mass of the dmp should not enter the result
at all in virtue of the equivalence principle. Thus, we arrive at
\begin{equation} \label{av}
<v \,\sigma> \,\sim \sqrt{54\pi}\;\;\frac{k^2\,m_p\,M}{u^3}\,,
\end{equation}
or
\begin{equation} \label{si}
\int_0^\infty dv \,v^3\, \sigma \,\sim \pi\, k^2\, m_p\,M\,;
\end{equation}
here $k$ is the Newton gravitation constant; an extra power of
$\pi$, inserted into these expressions, is perhaps inherent in
$\sigma$. Since the capture would be impossible if the planet were
not bound to the Sun, it is only natural that the result is
proportional to the corresponding effective "coupling constant"
$k\, m_p\, M$. One more power of $k$ corresponds to the
gravitational interaction of the dark matter particle. The final
estimate for the captured mass is
\begin{equation} \label{de1}
\mu_p \sim \rho_g T \sqrt{54\pi}\;\;k^2\,m_p\,M/u^3\,.
\end{equation}
For the Earth it constitutes
\begin{equation} \label{E}
\mu_E \sim 4\cdot10^{18}\,{\rm g}\,.
\end{equation}

\section{Capture cross-section}

By the same dimensional reasons (and in the complete
correspondence with formula (\ref{si})), the total capture
cross-section for the Earth should look as follows:
\begin{equation} \label{si1}
\sigma \sim \,\pi \,k^2\,m_E\,M\,/\tilde{v}^4\,,
\end{equation}
where $m_E$ is the mass of the Earth, and $\tilde{v}$ is some
velocity which can be estimated as follows. It is natural to
assume that the capture of dm particles occurs when they are close
to the Earth, i.e. at the distances $\sim r_E$ from the Sun. As
natural are the following assumptions: 1) the initial velocities
of the captured dmp's exceed only slightly the parabolic one
$v_{\rm par}$ ($v_{\rm par}^2 = 2kM/r_E$); 2) their final velocities are
only slightly less than $v_{par}$. To our accuracy, here we omit
the factor of 2 in the definition of $v_{\rm par}^2$, and thus put
$\tilde{v}^2 \sim v_E^2 = kM/r_E$ ($v_E = 30$ km/s is the velocity
of the Earth). Thus, the capture cross-section is
\beq\label{si2}
\sigma \,\sim \,\pi \,k^2\,m_E\,M\,/v_E^4\,.
\eeq
This formula can be also conveniently rewritten as
\beq\label{si3}
\sigma \,\sim \,\pi \,r_E^2\,(m_E/M)\,.
\eeq

Let us note here that the impact parameter corresponding to
formula (\ref{si3}), i.e. the typical distance between a dmp and
the Earth crucial for the capture, is
\beq\label{rho}
r_{\rm{imp}} \,\sim \,r_E\,(m_E/M)^{1/2}\,\ll r_E\,.
\eeq
In fact, this impact parameter corresponds to the distance at which
the attraction to the Earth equals the attraction to the Sun, i.e. where
\beq
km/r^2 > kM/r_E^2\, \;\;(r\ll r_E)\,.
\eeq.

Up to now, in all relevant formulae, (\ref{si}), (\ref{si1}),
(\ref{si2}), we dealt with the capture cross-section averaged
over the directions of the dmp velocity $\bv$. However, this
cross-section depends essentially on the mutual orientation of
$\bv$ and $\bv_E$. Certainly, it is maximum when these velocities
are parallel and as close as possible by modulus. Besides, the
impact parameter $r_{\rm{imp}}$ of the collision is much less than
the radius $r_E$ of the Earth orbit (see (\ref{rho})), and thus within
the distances $\simeq r_{\rm{imp}}$ both the Earth and dmp trajectories
can be treated as rectilinear. Therefore,
it looks quite natural to identify $\tilde{v}$ in (\ref{si1}) with
the relative velocity $|\bv -\bv_E|$ of the dmp and the Earth, i.e.
to generalize formula (\ref{si2}) as follows:
\beq \label{si4}
d\sigma \sim \,\frac{k^2\,m_p\,M}{(\bv -\bv_E)^4}\;\frac{1}{4} \,d\Omega
\eeq
(factor 1/4 is introduced here for correspondence with
factor $\pi$ in (\ref{si2}): $(1/4)\int d\Omega = \pi$).

Thus derived total cross-section is
\beq \label{si5}
\sigma \sim \,\frac{1}{4}\int d\Omega\;\frac{k^2\,m_p\,M}{(\bv -\bv_E)^4} \,
=\,\frac{\pi\,k^2\,m_p\,M}{(v^2 - v_E^2)^2}\,.
\eeq

Clearly, it is the particles moving initially with the velocities
only slightly above the parabolic one $\sqrt{2}\,v_E = 42$ km/s
that are captured predominantly, and thus, with $v =
\sqrt{2}\,v_E$, cross-sections (\ref{si2}) and (\ref{si5})
practically coincide.

On the other hand, it follows from (\ref{si5}) that in the vicinity of
the Earth the captured particles move with respect to it with the
velocities close to $(\sqrt{2}-1)v_p \simeq$ 12 km/s.

\section{Space distribution of captured dark matter}

The captured dmp's had initial trajectories predominantly close to
parabolas focussed at the Sun, and the velocities of these dmp's
change only slightly as a result of scattering. Therefore, their
trajectories become elongate ellipses with large semimajor axes,
still focussed at the Sun. The ratio of their maximum $r_{\rm
max}$ and minimum $r_{\rm min}$ distances from the Sun is
\cite{ll}
\beq
\frac{r_{\rm max}}{r_{\rm min}} = \frac{1 + e}{1 - e}\,,
\eeq
where $e$ is the eccentricity of the trajectory. In our case, as a
result of the capture, the eccentricity changes from $1+\ep_1$ to
$1-\ep_2$, where $\ep_{1,2} \ll 1$. This loss of eccentricity is
due to the gravitational perturbation by the Earth, and therefore
is proportional to $m_E$. On the other hand, $r_{\rm min}$ is
close to the radius $r_E$ of the Earth orbit. Thus, for
dimensional reasons, we arrive at \cite{khs}
\beq
r_{\rm max}\sim r_E \,(M/m_E)\,.
\eeq
Let us note here that the analogous estimate for the case of
Jupiter complies qualitatively with the results of corresponding
numerical calculations presented in \cite{pe}.

Obviously, the semimajor axis $a_{\rm dmp}$ of the trajectory of a
captured dmp is on the same order of magnitude as $r_{\rm max}$.
Then, the time spent by a dmp, with the characteristic velocity
close to $v_E$ and at the distance from the Sun close to $r_E$, is
comparable to the orbital period of the Earth $\;T_E = 1$ year.
Besides, the orbital period $T$ is related to the semimajor axis
$a$ as follows~\cite{ll}: $T \sim a^{3/2}$. Thus, we arrive at the
following estimate for the orbital period of the captured
dmp\footnote{In the case of the Earth, this orbital period is
huge, $\sim 10^8$ years. Still, it is much less than the lifetime
of the SS, $\sim 5 \cdot 10^9$ years.}:
\beq\label{td}
T_{\rm dmp}\sim T_E \,(M/m_E)^{3/2}\,.
\eeq
In other words, the relative time spent by a dmp at the distances
$\sim r_E$ from the Earth can be estimated as $(m_E/M)^{3/2}$.
Moreover, the typical distances from the Earth at which a dmp can
be captured, should be less than the impact parameter
$r_{\rm{imp}} \,\sim \,r_E\,(m_E/M)^{1/2}$ (see (\ref{rho})).
Thus, the relative time spent by a dmp sufficiently close to the
Earth to be captured, can be estimated as $(m_E/M)^2$.

With the impact parameter (\ref{rho}), the corresponding volume
$V$, centered at the Earth and crucial for the capture, can be estimated as
\beq\label{vol}
V \sim \frac{4\pi}{3}\,r_{\rm{imp}}^3 \,\sim
\frac{4\pi}{3}\,r_E^3\,(m_E/M)^{3/2}\,\ll \frac{4\pi}{3}\,r_E^3.
\eeq
Let us combine formula (\ref{E}) for the total captured mass with
the effective volume (\ref{vol}) occupied by this mass and with
the estimate $(m_E/M)^2$ for the relative time spent by a dmp
within the impact parameter (\ref{rho}) with respect to the Earth.
In this way we arrive at the following estimate for the density of
dark matter, captured by the SS, in the vicinity of the Earth:
\beq\label{dens}
\rho_E\sim 5 \cdot 10^{-25}\;\,{\rm g/cm^3}\,.
\eeq
This estimate practically coincides with the value (1) for the
galactic dm density.

In fact, the result ({\ref{dens}}) for the density of the captured dm,
as well as the estimates ({\ref{de1}}) and ({\ref{E}}) for its total mass, should be
considered as upper limits only, since we have
neglected therein the inverse process, that of the ejection of the
captured dm from the SS. The characteristic
time of the inverse process is not exactly clear now. Therefore, it cannot
be excluded that it is comparable to, or even larger than, the
lifetime $T$ of the SS \cite{khs}. In this case our estimates are valid.

If this is the case indeed, then the dm around the Earth consists essentially
of two components with comparable densities. In line with the common component
with the typical velocity around $u\sim 220$ km/s, there is one more, with the
velocity relative to the Earth $\gsim 12$ km/s.

\bigskip

{\bf Acknowledgements.} I am grateful to V.V. Sokolov for useful discussions.
The work was supported by the Russian Foundation for Basic Research through Grant No.
08-02-00960-a.

\end{document}